\documentclass[final,5p,times,twocolumn]{elsarticle}
\usepackage{graphicx}
\usepackage{amssymb}
\usepackage{amsthm}

\journal{Physics Letters A}

\begin{document}

\begin{frontmatter}

\title{
Zipf's law from a Fisher variational-principle}

\author[a1]{A. Hernando}
\ead{alberto@ecm.ub.es}
\author[a1]{D. Puigdom\`enech}
\ead{puigdomenech@ecm.ub.es}
\author[a2]{D. Villuendas}
\ead{diego@ffn.ub.es}
\author[a3]{C. Vesperinas}
\ead{cristina.vesperinas@sogeti.com}
\author[a4]{A. Plastino}
\ead{plastino@fisica.unlp.edu.ar}
\address[a1]{Departament ECM, Facultat de F\'{\i}sica,
Universitat de Barcelona. Diagonal 647,
08028 Barcelona, Spain}
\address[a2]{Departament FFN, Facultat de F\'isica, Universitat de Barcelona,
Diagonal 647, 08028 Barcelona, Spain}
\address[a3]{Sogeti Espa\~na, WTCAP 2, Pla\c ca de la Pau s/n, 08940
Cornell\`a, Spain}
\address[a4]{National University La Plata, IFCP-CCT-CONICET, C.C. 727, 1900 La Plata, Argentina}

\begin{abstract}

Zipf's law is shown to arise as the variational solution of a
problem formulated in Fisher's terms. An appropriate minimization
process involving Fisher information and scale-invariance yields
this universal rank distribution. As an example we show that the
number of citations found in the most referenced physics journals follows this law.

\end{abstract}


\end{frontmatter}


\section{Introduction}
\label{p1}

This work discusses the application of Fisher's information measure
to some scale-invariant phenomena. We thus begin our considerations with
a brief review of the pertinent ingredients.

\subsection{Scale-invariant phenomena}

The study of scale-invariant phenomena has unravelled interesting
and somewhat unexpected behaviours in systems belonging to
disciplines of different nature, from physical and biological to
technological and social sciences~\cite{uno}. Indeed, empirical data
from percolation theory and nuclear multifragmentation~\cite{perco}
reflect scale-invariant behaviour, and so do the abundances of genes
in various organisms and tissues~\cite{furu}, the frequency of words
in natural languages~\cite{zip}, scientific collaboration
networks~\cite{cites}, the Internet traffic~\cite{net1}, Linux
packages links~\cite{linux}, as well as electoral
results~\cite{elec1}, urban agglomerations~\cite{ciudad,ciudad2} and
firm sizes all over the world~\cite{firms}.

The common feature in these systems is the lack of a characteristic
size, length or frequency for an observable $k$ at study. This lack
generally leads to a power law distribution $p(k)$, valid in most of
the domain of definition of $k$,
\begin{equation}\label{eq1}
p(k)\sim1/k^{1+\gamma},
\end{equation}
with $\gamma\geq0$. Special attention has been paid to the class of
universality defined by $\gamma=1$, which corresponds to Zipf's law
in the cumulative distribution or the rank-size
distribution~\cite{perco,furu,zip,net1,linux,ciudad,ciudad2,firms,citis}.
Recently, Maillart et al.~\cite{linux} have studied the evolution of
the number of links to open source software projects in Linux
packages, and have found that the link distribution follows Zipf's
law as a consequence of stochastic proportional growth. In its
simplest formulation, the  stochastic proportional growth
 model, or namely the geometric Brownian motion,  assumes the growth of an
element of the system to be proportional to its size $k$, and to be
governed by a stochastic Wiener process. The class $\gamma=1$
emerges from the condition of stationarity, i.e., when the system
reaches a dynamic equilibrium~\cite{citis}.
Together with geometric Brownian motion, there is a variety of
models arising in different fields that yield Zipf's law and other
power laws on a case-by-case basis~\cite{ciudad,ciudad2,citis,mod1,exp},
as preferential attachment~\cite{net1} and competitive cluster
growth~\cite{ccg} in complex networks, used to explain many of the scale-free properties
of social, technological and biological networks.

\subsection{Fisher's information measure}

Much effort has recently been devoted to  Fisher's information measure (FIM), usually denoted as $I$. The work of Frieden
and
co-workers~\cite{libro,    Fri89,Fri92,Fri94,Fri93,FH94,NF94,Fri90,FS95,Fri91},
Silver~\cite{ST92}, and Plastino et
al.~\cite{PPMK96,PP96,PPM97,PPM97pla}, among many others,  has shed much light upon the
manifold physical applications of $I$. As a small sample we mention that Frieden and Soffer have shown that FIM
provides a powerful variational principle, called EPI (extreme physical information) that yields the
canonical Lagrangians of theoretical physics~\cite{FS95}.
Additionally, $I$ has been proved to characterize an ��arrow of time��
with reference to the celebrated Fokker-Planck equation~\cite{PP96}.
Moreover, there exist interesting relations that connect FIM and the
relative Shannon information measure invented by
Kullback~\cite{Kul59,RCP97}. These can be shown to have some
bearing on the time evolution of arbitrary systems governed by quite
general continuity equations~\cite{PPM97,PPM97pla}. Additionally, a
rather general $I$-based H theorem has recently been
proved~\cite{PCP98,Fri98}. As for Hamiltonian
systems~\cite{H}, EPI allows to describe the behaviour of complex systems, as the
allometric or power laws found in biological
sciences~\cite{allometric}. The pertinent list could be extended quite a bit. $I$ is then an important quantity, 
involved in many aspects of the theoretical description of nature.

For our present purposes it is of the essence to mention that Frieden et al.~\cite{fisher2} have also shown that equilibrium and
non-equilibrium thermodynamics can be derived from a principle of
minimum Fisher information, with suitable constraints (MFI). Here $I$ is specialized
to the particular but important case of \emph{translation
families}, i.e., distribution functions whose form does not change
under translational transformations. In this case, Fisher measure
becomes \emph{shift-invariant}.  It is shown
in~\cite{fisher2} than such minimizing of Fisher's measure leads to a
Schr\"odinger-like equation for the probability amplitude, where the
ground state describes equilibrium physics and the excited states
account for non-equilibrium situations.

\subsection{Goals and motivation}

Scale-invariant phenomena are generally addressed by appeal to
ad-hoc models (see the references citing in 1.1). In spite of the
success of these models, the intrinsic complexity involved therein makes their
study at a macroscopic level a rather  difficult task. One sorely misses a general formulation
of the thermodynamics of scale-invariant physics, which  is not quite established
yet. It is our goal here to show, in such a vein, that minimization of Fisher
information provides a unifying framework that allows these phenomena to be understood
 as arising from an MFI variational principle, entirely analogous to how termodynamics is generated in [34].

\section{Minimum Fisher Information approach (MFI)}
\label{p2}

The Fisher information measure $I$ for a system described by a set
of coordinates $\mathbf{q}$ and physical parameters
$\mathbf{\theta}$, has the form~\cite{Fri98}
\begin{equation}\label{fish}
I(F)=\int_\Omega
d\mathbf{q}F(\mathbf{q}|\mathbf{\theta})\sum_{ij}c_{ij}\frac{\partial}{\partial\theta_i}\ln
F(\mathbf{q}|\mathbf{\theta})\frac{\partial}{\partial\theta_j}\ln
F(\mathbf{q}|\mathbf{\theta}),
\end{equation}
where $F(\mathbf{q}|\mathbf{\theta})$ is the density distribution in
a configuration space ($\mathbf{q}$) of volume $\Omega$ conditioned
by the physical parameters ($\mathbf{\theta}$). The constants
$c_{ij}$ account for dimensionality, and take the form
$c_{ij}=c_i\delta_{ij}$ if $q_i$ and $q_j$ are uncorrelated.
The equilibrium state of the system minimizes $I$ subject to prior
conditions, like the normalization of $F$ or any constraint on the
mean value of an observable $\langle A_i \rangle$~\cite{fisher2}.
The MFI is then written as a variation problem of the form
\begin{equation}
\delta\left\{I(F)-\sum_i\mu_i\langle A_i \rangle\right\}=0,
\end{equation}
where $\mu_i$ are appropriate Lagrange multipliers.

\subsection{One-dimensional system with discrete coordinate}

Because of the nature of the systems to be addressed we consider now
a one-dimensional system with a physical parameter $\theta$ and a
discrete coordinate $k=k_1,k_2,\ldots,k_i,\ldots$ where
$k_{i+1}-k_i=\Delta k$ for a certain value of the interval $\Delta
k$. This scenario arises, for instance, in the case of nuclear
multifragmentation~\cite{perco}, the abundances of
genes~\cite{furu}, the frequency of words~\cite{zip}, scientific
collaboration networks~\cite{cites}, the Internet
traffic~\cite{net1}, Linux packages links~\cite{linux}, electoral
results~\cite{elec1}, urban agglomerations~\cite{ciudad,ciudad2},
firm sizes~\cite{firms}, etc.

In the continuous limit ($\Delta k\rightarrow dk$), the Fisher
information measure is cast as
\begin{equation}
I(F)=c_k\int_{k_1}^\infty dk
F(k|\theta)\left|\frac{\partial}{\partial\theta}\ln
F(k|\theta)\right|^2.
\end{equation}
Instead of using translation invariance \`a la Frieden-Soffer~\cite{FS95}, 
we will appeal to scaling invariance~\cite{sym} so that
we can anticipate some new physics. All members of the family
$F(k/\theta)$ possess identical shape ---there are no characteristic
size, length or frequency for the observable $k$--- namely
$dkF(k/\theta)=dk'F(k')$ under the transformation $k'=k/\theta$.

To deal with this new symmetry it is convenient to change to the new
coordinate $u=\ln k$ and parameter $\Theta=\ln \theta$. Why? Because then
the scale invariance becomes again translational invariance, and we are entitled  to use one essential result of [34], namely, that MFI leads  to a Schroedinger-like equation. Note that  
the new coordinate $u'=\ln k'$ transforms as $u'=u-\Theta$. Defining
$f(u)=F(e^u)$ and taking into account the fact that the Jacobian of
the transformation is $|dx/du|=e^u$ and
$\partial/\partial\theta=e^{-\Theta}\partial/\partial\Theta$, the
Fisher information measure acquires now the form
\begin{equation}\label{f2}
I(F)=c_k e^{-2\Theta}\int_{u_1}^\infty
du~e^{u}f(u)\left|\frac{\partial\ln f(u)}{\partial u}\right|^2,
\end{equation}
where $u_1=\ln k_1$, and the factor $e^{-2\Theta}$ guaranties the
invariance of the associated Cramer-Rao inequality as shown in~\cite{sym}.

For reasons that will become apparent below, we will apply the
MFI without any constraint. This is tantamount to posing no bound to
the physical ``sizes'' that characterize the system. The
extremization of Fisher information with no constraints ($\mu_i=0$)
is written as
\begin{equation}\label{vargII}
\delta\left\{\int_{u_1}^\infty du~e^{u}f(u)\left|\frac{\partial\ln
f(u)}{\partial u}\right|^2\right\}=0.
\end{equation}
Introducing $f(u)=e^{-u}\Psi^2(u)$, and varying with respect to
$\Psi$ and $\partial\Psi/\partial u$ as in~\cite{fisher2} one is
easily led to a (real) Schr\"odinger-like equation of the form
\begin{equation}\label{schr}
\left[-4\frac{\partial^2}{\partial u^2}+1\right]\Psi(u)=0.
\end{equation}
Notice that the lack of normalization constraints implies zero
eigenvalue, since the Lagrange multiplier associated with the
normalization is the energy eigenvalue~\cite{fisher2}. At this point
we introduce boundary conditions to guaranty convergence of the
Fisher measure (\ref{f2}) and thus compensate for the lack of
constraints in (\ref{vargII}). We impose $\lim_{u\rightarrow\infty}\Psi(u)=0$ and
$\Psi(u_1)=\sqrt{N}$, where $N$ is an dimensionless constant
the meaning of which will become clear later. 
The
solution to (\ref{schr}) with these boundary conditions is
$\Psi(u)=\sqrt{N}e^{-(u-u_1)/2}$, which leads to
$f(u)=Ne^{-(2u-u_1)}$ and to the density distribution
\begin{equation}\label{zip}
F(k)dk=N\frac{k_1}{k^2}dk
\end{equation}
with $N=1$ for a density normalized to unity. This distribution is
just the Zipf's law (universal class $\gamma=1$) of
Refs.~\cite{perco,furu,zip,net1,linux,ciudad,firms,citis}. This
result is remarkable: \emph{Zipf's law has been here derived from
first principles}.

\section{Applications}

A common representation of empirical data is the so-called rank-plot
or Zipf plot~\cite{zip,ciudad2,mod1}, where the $j$th element of the
system is represented by its size, length or frequency $k_j$ against
its rank, sorted from the largest to the smallest one. This process
just renders the inverse function of the ensuing cumulative
distribution, normalized to the number of elements. We call $r$ the
rank that ranges from 1 to $N$. Thus, the constant $N$ arising 
from the boundary conditions is the total number
of elements considered in building up the distribution (\ref{zip}),
as will be illustrated in the examples bellow. This
rank-distribution takes the form
\begin{equation}\label{ziprank}
k(r)=N\frac{k_1}{r}
\end{equation}
which yields a straight line in a logarithmic representation with
slope $-1$.

In Fig.~1a we depict the known behavior~\cite{citis} of the rank
size distribution for the top 100 largest cities of the United
States~\cite{usa}, which shows a slope near $-1$ ($\gamma=1$) in
the logarithmic representation of the rank-plot.

We have also studied the system formed by the most referenced physics
journals~\cite{isi}, using their total number of cites as coordinate
$k$. If a journal receives more cites due to its popularity, it
becomes even more popular and, therefore,  receives still more cites, etc.
Under such conditions, proportional growth and scale invariance are
expected, as we depict in Fig.~1b, where the slope's value can be regarded as illustrating the
universality of the underlying law.

\begin{figure}[t!]
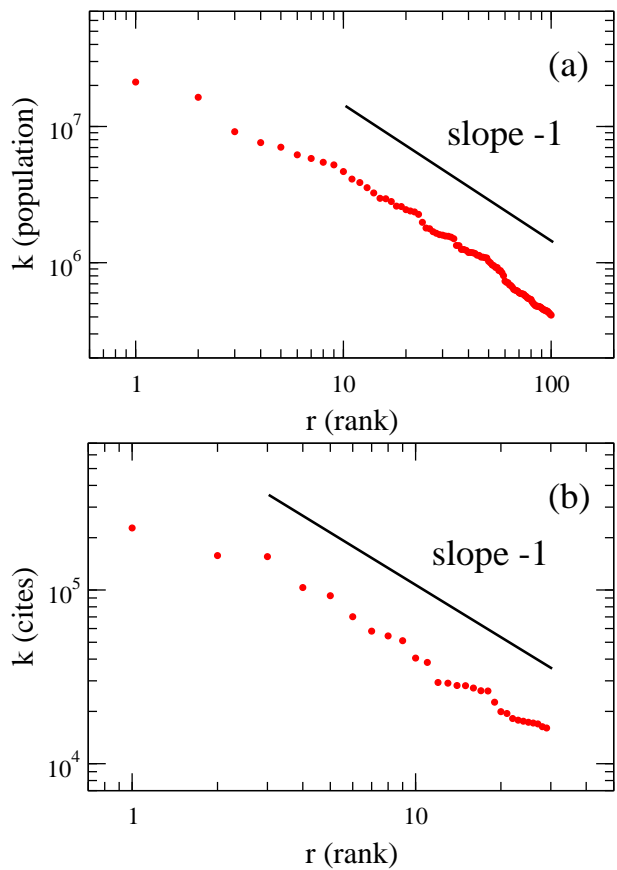

\includegraphics[width=0.9\linewidth,clip=true]{fig1a.eps}\\
\includegraphics[width=0.9\linewidth,clip=true]{fig1b.eps}
\caption{a. Rank-plot of the 100 largest cities of the United
States, from most-populated to less-populated, in logarithm
scale. b. Rank-plot of the total number of cites of the 30 most
cited physics journals, from most-cited to less-cited, in logarithm
scale.}
\end{figure}

\section{Conclusions}

We have here shown that Zipf's law results from the scaling
invariance of the Crammer-Rao inequality derived in [35]. This entails that the
relevant probability distribution, usually called the
rank-distribution, has to be size-invariant. Consequently,  it should
be derivable from a minimization process in which Fisher's
information measure is the protagonist. No constraints are needed in
the concomitant variational problem because, a priori, our sizes have no upper
bound. A physical analogy is the non-normalizability of plane waves.
The universal character of our demonstration thus resides in the 
universal form to be minimized (Fisher's), with no constraints.

\section*{Acknowledgments}

We would like to thank M. Barranco, R. Frieden, and B. H. Soffer for
useful discussions. This work has been partially performed under
grant FIS2008-00421/FIS from DGI, Spain (FEDER).


\begin{thebibliography}{99}

\bibitem{uno} Fractals in Physics: Essays in Honour of Benoit B Mandelbrot, edited by A. Aharony and J. Feder, Proceedings
of a Conference in honor of B.B. Mandelbrot on His 65th in Vence, France, North Holland, Amsterdam, 1989.

\bibitem{perco} K. Paech, W. Bauer, and S. Pratt , { Phys. Rev. C} {76} (2007), p. 054603; X. Campi and H. Krivine, { Phys. Rev. C} {72} (2005) 057602;
Y. G. Ma et al., { Phys. Rev. C} {71} (2005) 054606.

\bibitem{furu} C. Furusawa and K. Kaneko, { Phys. Rev. Lett.} {90} (2003) 088102.

\bibitem{zip} G. K. Zipf, Human Behavior and the Principle of Least Effort,
Addison-Wesley Press, Cambridge, Mass., 1949; I. Kanter and D. A.
Kessler, { Phys. Rev. Lett.} {74} (1995) 4559.

\bibitem{cites} M. E. J. Newman, { Phys. Rev. E} {64} (2001) 016131.

\bibitem{net1} A.-L. Barabasi and R. Albert, { Rev. Mod. Phys.} {74} (2002) 47.

\bibitem{linux} T. Maillart, D. Sornette, S. Spaeth, and G. von Krogh, { Phys. Rev. Lett.} {101} (2008) 218701.

\bibitem{elec1} R. N. Costa Filho, M. P. Almeida, J. S. Andrade, and J. E. Moreira, { Phys. Rev. E} {60} (1999) 1067.

\bibitem{ciudad} L. C. Malacarne, R. S. Mendes, and E. K. Lenzi, { Phys. Rev. E} {65} (2001) 017106.

\bibitem{ciudad2} M. Marsili and Yi-Cheng Zhang, { Phys. Rev. Lett.} {80} (1998) 2741.

\bibitem{firms} R. L. Axtell, { Science} {293} (2001) 1818.

\bibitem{citis} X. Gabaix, { Quarterly Journal of Economics} {114} (1999) 739.

\bibitem{mod1} K. E. Kechedzhi, O. V. Usatenko, and V. A. Yampol'skii, { Phys. Rev. E} {72}, 046138
(2005); S. Ree, { Phys. Rev. E} {73} (2006) 026115.

\bibitem{exp} W. J. Reed, and B. D. Hughes, { Phys. Rev. E} {66} (2002) 067103.

\bibitem{ccg} A. A. Moreira, D. R. Paula, R. N. Costa Filho, and J. S. Andrade, { Phys. Rev. E} {73} (2006) 065101(R).

\bibitem{libro} B. R. Frieden, Science from Fisher Information, Cambridge University Press, Cambridge, England, 2004.

\bibitem{Fri89} B. R. Frieden, { Am. J. Phys.} { 57} (1989) 1004.

\bibitem{Fri92} B. R. Frieden, { Phys. Lett. A} { 169} (1992) 123.

\bibitem{Fri94} B. R Frieden, in Advances in Imaging and Electron Physics, edited by
P. W. Hawkes, Academic, New York, 1994, Vol. 90, pp. 123�204.

\bibitem{Fri93} B. R. Frieden, { Physica A} { 198} (1993) 262.

\bibitem{FH94} B. R. Frieden and R. J. Hughes, { Phys. Rev. E} { 49} (1994) 2644.

\bibitem{NF94} B. Nikolov and B. R. Frieden, { Phys. Rev. E} { 49} (1994) 4815.

\bibitem{Fri90} B. R. Frieden, { Phys. Rev. A} { 41} (1990) 4265.

\bibitem{FS95} B. R. Frieden and B. H. Soffer, { Phys. Rev. E} { 52} (1995) 2274.

\bibitem{Fri91} B. R. Frieden, { Found. Phys.} { 21} (1991) 757.

\bibitem{ST92} R. N. Silver, in E. T. Jaynes: Physics and Probability, edited by W.
T. Grandy, Jr. and P. W. Milonni, Cambridge University Press, Cambridge, England, 1992.

\bibitem{PPMK96} A. Plastino, A. R. Plastino, H. G. Miller, and F. C. Khana, { Phys. Lett. A} { 221} (1996) 29.

\bibitem{PP96} A. R. Plastino and A. Plastino, { Phys. Rev. E} { 54} (1996) 4423.

\bibitem{PPM97} A. R. Plastino, A. Plastino, and H. G. Miller, { Phys. Rev. E} { 56} (1997) 3927.

\bibitem{PPM97pla} A. Plastino, A. R. Plastino, and H. G. Miller, { Phys. Lett. A} { 235} (1997)
129.

\bibitem{Kul59} S. Kullback, Information Theory and Statistics, Wiley, New York, 1959.

\bibitem{RCP97} M. Ravicule, M. Casas, and A. Plastino, { Phys. Rev. A} { 55} (1997) 1695.

\bibitem{PCP98} A. R. Plastino, M. Casas, and A. Plastino, { Phys. Lett. A} { 246} (1998) 498.

\bibitem{Fri98} B. R. Frieden, Physics from Fisher Information, Cambridge University
Press, Cambridge, England, 1998.

\bibitem{H} F. Pennini and A. Plastino, { Phys. Lett. A} {349} (2006) 15.

\bibitem{allometric} B. R. Frieden and R. A. Gatenby, { Phys. Rev. E} {72} (2005) 036101.

\bibitem{fisher2} B. R. Frieden, A. Plastino, A. R. Plastino, and B. H. Soffer,
Phys. Rev. E {60} (1999) 48; {66} (2002) 046128.

\bibitem{sym} F. Pennini, A. Plastino, B. H. Soffer, and C. Vignat, { Phys. Let. A} {373} (2009) 817.

\bibitem{usa} Census bureau website, Government of the USA, www.census.gov.

\bibitem{isi} Journal Citation Reports (JCR), Thomson Reuters, 2007.



\end{thebibliography}
\end{document}